\documentclass[runningheads]{llncs}
\usepackage{graphicx}
\usepackage{microtype}
\usepackage{booktabs} 
\usepackage{amsfonts}
\usepackage{epstopdf}
\usepackage{enumerate}
\usepackage{longtable}
\usepackage{url}
\usepackage{CJK}
\usepackage{verbatim}
\usepackage{multirow} 
\usepackage{pgfplots}
\usepackage{pgfplotstable}
\usepackage{wrapfig}
\usetikzlibrary{positioning}
\usepackage{multirow}
\usepackage{diagbox}
\usepackage{booktabs}
\usepackage{colortbl}
\usepackage{caption}
\usepackage{longtable}
\usepackage{url}
\usepackage{cite}
\usepackage{amsmath,amssymb,amsfonts}
\usepackage{textcomp}
\usepackage{appendix}
\usepackage{color}
 \usepackage{multirow}
\usepackage{tikz,xcolor}
\usetikzlibrary{positioning,automata,shadows}
 \makeatletter
    \newif\if@restonecol
    \makeatother

    \usepackage[linesnumbered,ruled,vlined]{algorithm2e}
    \usepackage{algpseudocode}
    \usepackage{amsmath}

\definecolor{liyt_red}{RGB}{254,117,109}
\definecolor{liyt_green}{RGB}{108,206,157}
\definecolor{liyt_blue}{RGB}{102,188,243}
\definecolor{liyt_gray}{RGB}{166,170,179}
\begin{document}
\title{Learning Restricted Regular Expressions with Interleaving}
\titlerunning{Learning Restricted Regular Expressions with Interleaving}
%
%
\author{Chunmei Dong\inst{1,2} \and Yeting Li\inst{1,2} \and Haiming Chen\inst{1}}
\authorrunning{C. Dong et al.}
%
\institute{University of Chinese Academy of Sciences, Beijing, China\\  \and
State Key Laboratory of Computer Science, Institute of Software, Chinese Academy of Sciences, Beijing, China\\
\email{\{dongcm,liyt,chm\}@ios.ac.cn}}
\maketitle

\begin{abstract}
The advantages for the presence of an XML schema for XML documents are numerous. However, many XML documents in practice are not accompanied by a schema or by a valid schema. Relax NG is a popular and powerful schema language, which supports the unconstrained interleaving operator. Focusing on the inference of Relax NG, we propose a new subclass of regular expressions with interleaving and design a polynomial inference algorithm. Then we conducted a series of experiments based on large-scale real data and on three XML data corpora, and experimental results show that our subclass has a better practicality than previous ones, and the regular expressions inferred by our algorithm are more precise.
\end{abstract}

\section{Introduction}
As a main file format for data exchange, the eXtensible Markup Language (XML) has been widely used on the Web \cite{Abiteboul2000Data}. XML schemas define the structure constraints of XML documents. The advantages by the presence of an XML schema for XML documents are numerous, such as for data processing, automatic data integration, static analysis of transformations and so on \cite{Benedikt2008XPath,Che2006Query,Koch2004Schema,Manolescu2001Answering,Martens2003Typechecking,Martens2004Frontiers,Papakonstantinou2000DTD}.
However, many XML documents are not accompanied by a (valid) schema in practice. Research in 2013 showed that only $24.8\%$ XML documents available on the Web were accompanied with corresponding schemas, of which the proportion of valid ones was only $8.9\%$ \cite{grijzenhout2013quality}. Therefore, it has become an urgent problem to infer a suitable XML schema for given XML documents.

Document Type Definition (DTD), XML Schema Definition (XSD) and Relax NG are three popular XML schema languages.
Among them  Relax NG is more powerful than both DTD and XSD ¡ªdue to its expressive power \cite{DBLP:journals/siamcomp/PaigeT87}.
Relax NG schemas support the interleaving operator and allow the interleaving to be mixed with other operators,
which can make the schemas succinct. Furthermore, the interleaving has been used in many applications. For example, it is used in the schema language ShEx for RDF \cite{DBLP:journals/corr/BonevaGPS15,DBLP:conf/icdt/StaworkoBGHPS15}, and is necessary in solving many problems \cite{DBLP:conf/icde/LiG15,DBLP:conf/kdd/FowkesS16,DBLP:conf/chi/KinHDA12}. On the other hand, presently researches on interleaving are quite insufficient (for instance, see below). Therefore we concentrate on the study of interleaving, and focus on the inference of Relax NG in this paper. Notice that the results can also be applied to some other applications.

Actually, the major task of schema inference can be reduced to learning regular expressions from a set of given samples \cite{Bex2006Inference,Bex2007Inferring,DBLP:journals/datamine/GarofalakisGRSS03}. Gold proposed a classical language learning model (\textit{learning in the limit or explanatory learning}) and pointed out that the class of regular expressions could not be identifiable from positive samples only \cite{Gold1967Language}. Therefore, researches have focused on learning restricted subclasses of regular expressions \cite{Min2003Efficient}.

For the inference of regular expressions, Bex et al. proposed two subclasses: \textit{single occurrence regular expressions} (SOREs) and \textit{chain regular expressions} (CHAREs), and gave their inference algorithms \textit{RWR} and \textit{CRX} \cite{Bex2006Inference,Bex2010Inference}. Freydenberger et al. gave two more efficient algorithms \textit{Soa2Sore} and \textit{Soa2Chare} for the above classes. Kim et al. developed an inference system using hedge grammars
for learning Relax NG \cite{DBLP:conf/lata/KimKH16}. However, all of the above work are based on standard regular expressions, which do not support interleaving. Nevertheless, interleaving is vital since there may be no order constraint among siblings in data-centric applications \cite{abiteboul2015highly}, and it has been proved that regular expressions with interleaving are double exponentially more succinct than standard regular expressions \cite{Gelade2008Succinctness}.

 With regard to the regular expressions with interleaving, Ghelli et al. proposed a restricted subclass called ``conflict-free REs'' supporting interleaving \cite{DBLP:conf/dbpl/GhelliCS07}, where the subclass requires no symbol appears twice and repetition is only applied to single symbols, and no inference algorithm is provided. Ciucanu and Staworko proposed two subclasses called disjunctive multiplicity expressions (DME) and disjunction-free multiplicity expressions (ME) \cite{DBLP:conf/webdb/BonevaCS13}, which support unordered concatenation (a weaker form of
interleaving), and disallow concatenation within siblings. The inference algorithm of DME is discussed in \cite{DBLP:journals/corr/CiucanuS13}. Peng et al. \cite{peng2015discovering} proposed a subclass called the subset of regular expressions with interleaving (SIREs), and gave its inference algorithm based on the maximum independent set. 

In this paper, to address the problem of inferencing Relax NG, we propose a restricted subclass of regular expressions with interleaving, named as Improved Subclass of Regular Expressions with Interleaving (ISIREs), and SIREs is a subclass of ISIREs. We also develop its learning algorithm. The main contributions of this paper are listed as follows.
\begin{itemize}
  \item We propose a new subclass of regular expressions supporting interleaving, which helps the inference of RELAX NGs.
  \item We develop an inference algorithm InferISIRE to infer ISIREs and analyze its time complexity.
  \item We calculate the usage proportion of ISIREs and other popular subclasses based on the large-scale real data, and find that the proportion of ISIREs is the highest, which indicates ISIREs have a better practicality. Then, based on three XML data corpora, we compare the inferred results of InferISIRE with other inference algorithms. Experimental results show that the regular expressions inferred by InferISIRE are more precise.
\end{itemize}

The rest of this paper is organized as follows. Section $2$ presents the basic definitions. Section $3$ gives the inference algorithm InferISIRE. Section $4$ introduces the experiments. Conclusions are drawn in Section $5$.

\section{Preliminaries}
\begin{definition}
\textbf{Regular Expression with Interleaving}. Let $\Sigma$ be a finite alphabet. A string is a finite sequence of symbols
over $\Sigma$. The set of all finite strings over $\Sigma$ is denoted by $\Sigma^*$. The empty string is denoted by $\varepsilon$. A regular expression with interleaving over $\Sigma$ is defined inductively as follows: $\varepsilon$ or $a \in \Sigma$ is a regular expression, for regular expressions $E_1$ and $E_2$, the disjunction $E_1 | E_2$, the concatenation $E_1\cdot E_2$, the interleaving $E_1\&E_2$, or the Kleene-Star $E_1^*$ is also a regular expression. Usually we use $E_1E_2$ instead of $E_1\cdot E_2$ for readability. The length of a regular expression $E$, denoted by $|E|$, is the total number of alphabet symbols and operators occurring in $E$. The language generated by E is defined as follows: $L(\emptyset)=\emptyset$; $L(\varepsilon)=\{\varepsilon\}$; $L(a)=\{a\}$; $L(E_1^*)=L(E)^*$; $L(E_1E_2)=L(E_1)L(E_2)$; $L(E_1|E_2)=L(E_1)\cup L(E_2)$; $L(E_1\& E_2)=L(E_1)\& L(E_2)$. $E^?$ and $E^+$ are used as abbreviations of $E|\varepsilon$ and $EE^*$, respectively.
\end{definition}

Let $u=au_0,v=bv_0$, where $a,b\in \Sigma$ and $u,u_0,v,v_0\in \Sigma^*$. Then $u\& \varepsilon = \varepsilon \& u=\{u\}$; $u\&v=\{a(u_0\&v)\}\cup \{b(u\&v_0)\}$. For example, the string set generated by $a \& bc$ is $\{abc,bac,bca\}$.

Regular expressions with interleaving, in which each symbol occurs at most once, are called \textit{ISOREs} extended from \textit{SOREs} \cite{Bex2006Inference}.
\begin{definition}\textbf{Single Occurrence Automaton (SOA)} \cite{Bex2010Inference} Let $\Sigma$ be a finite alphabet, and let $src$ and $snk$ be distinct symbols that do not occur in $\Sigma$. A single occurrence automaton over $\Sigma$ is a finite directed graph $G$=$(V,D)$ such that
\vspace{-6pt}
\begin{itemize}
  \item $src,snk$$\in$$V$, and $V$$\subseteq$$\Sigma$$\cup$$\{src,snk\}$;
  \item $src$ has only outgoing edges, $snk$ has only incoming edges and every node $v$$\in$$V$ lies on a path from $src$ to $snk$.
\end{itemize}
\end{definition}

A \textit{generalized single occurrence automaton (generalized SOA)} over $\Sigma$ is defined as a directed graph in which each node $v$$\in$$V$$\setminus$$\{src,snk\}$ is an \textit{ISORE} and all nodes are pairwise \textit{ISOREs}.

For a given directed acyclic graph $G = (V,E)$ of an SOA, the level number of node $v$ is the longest path from
$src$ denoted by $ln(v)$, and $ln(src)$ is 0 \cite{Freydenberger2015Fast}. For a path $v_1$ to $v_2$ , if there exists node $v$ such that $ln(v_1)<ln(v)$ and $ln(v)<ln(v_2)$, then $ln(v)$ is a skip level \cite{Freydenberger2015Fast}.

\textit{Partial Order Relation (POR)} is a binary relation which is reflexive, dissymmetric and transitive. It is a collection of ordered pairs of elements over $\Sigma$. For a string $s=s_1s_2\cdot \cdot \cdot s_n$, $s_i \prec s_j$ means $s_i$ occurs before $s_j$. Using $POR$, we can construct the Constraint Set ($CS$) and Non-Constraint Set ($NCS$) for a set of given samples $S$. $POR(S)$ is denoted to represent the set of all partial orders obtained from each string $s \in S$. The formula is as follows.
\begin{itemize}
  \item $CS(S)=\{<a_i,a_j>|<a_i,a_j>$ $\in POR(S), and <a_j,a_i>$ $\in POR(S)\}$;
  \item $NCS(S)=\{<a_i,a_j>|<a_i,a_j>$ $\in POR(S), but <a_j,a_i>$ $\notin POR(S)\}$.
\end{itemize}
\begin{definition}
\textbf{Improved Subclass of Regular Expressions with Interleaving (ISIREs)} Let $\Sigma$ be a finite alphabet and $a$$\in$$\Sigma$, the improved subclass of regular expressions with interleaving
is defined by the following grammar: $S :: = TS|T$, $T :: = A\&T|A$, $A:: = \varepsilon|a|a^*|A A$.

Moreover, we require that every symbol $a$$\in$$\Sigma$ can occur at most once in the regular expression that belongs to ISIREs.
\end{definition}

By definition, it is obvious that ISIREs is a subclass of ISOREs and SIREs \cite{peng2015discovering} is a subclass of ISIREs. For example, $E_1=a\&b^*c\&de^?$ is an SIRE \cite{peng2015discovering} and also an ISIRE, $E_2=a^?(bc^?\&d^+e)f^*$ is an ISIRE but not an SIRE \cite{peng2015discovering}, and both $E_1$ and $E_2$ are ISOREs.

\section{Learning Algorithm for ISIREs}
In this section, we first introduce our learning algorithm InferISIRE, which infers an ISIRE for a set of given samples $S$. Then we analyse the time complexity of the algorithm. Finally, we show the learning process of InferISIRE through an example
and compare it with other algorithms.
\subsection{Algorithm Implementation and Analysis}
The algorithm InferISIRE uses a number of subroutines, and we introduce some of them as follows.
\begin{itemize}
  \item[$\bullet$] \textbf{cntOper($S$)}. $cntOper(S)$ is a function to calculate unary operator of every symbol in a set of strings $S$. The function returns a dictionary of pair $(a:opt)$, where $a$ is symbol in $S$ and $opt$ is an unary operator in $\{1,?,*,+\}$ , and the $opt$ is computed according to the occurences of $a$ in strings set $S$. For example, given the set of strings $S=\{aabd,abcd,bbcd\}$, $cntOper(s)=\{a:*;b:+;c:?;d:1\}$.
  \item[$\bullet$] \textbf{Filter($mis,consist\_tr$)}. For a maximum independent set (MIS) $mis$ and a non-constraint set $consist\_tr$, the function returns a subset of $consist\_tr$ consisting of the ordered pairs of elements over $mis$. That is, Filter($mis$,$consist\_tr$) \\=$\{<x,y>|<x,y>$ $\in consist\_tr$, and $x\in mis, y \in mis\}$;
  \item[$\bullet$] \textbf{contract($U, SubRE$)} \cite{Freydenberger2015Fast}. For a nodes set $U$ and a subexpression $SubRE$, the contract on an SOA modifies SOA such that all nodes of $U$ are contracted to a single vertex and labeled $SubRE$ (corresponding edges are moved).
\end{itemize}

The pseudo-code of the inference algorithm InferISIRE is shown in Algorithm \ref{alg1}, which outputs an ISIRE for an input set of strings. We illustrate the main procedures of  InferISIRE in the following.
\begin{enumerate}
  \item Traverse the samples $S$ to get the alphabet $\Sigma$, and get the dictionary $dictOper$ with the function $cntOper(S)$. This step is shown in line 1.
  \item We compute the non-constraint set $consist\_tr$ and constraint set $constraint\_tr$, 
      and construct an undirected graph $G$ using symbols in $\Sigma$ as nodes and ordered pairs in constraint set as edges. We then get the set $subGraphs$ consisting of all maximal connected subgraph of $G$ using function $connected\_subgraph(G)$. The lines 2-4 show the process.
  \item In line 6 and line 7, for each maximal connected subgraph of $G$, namely every $subgraph$ in the set $subGraphs$, we transform it to a subexpression with algorithm G2SubRE showed in Algorithm 2 (introduced later). We get the set $R$ of all subexpressions.
  \item We use $2T$-$INF$ \cite{Garcia2002Inference} to construct the SOA $\mathcal{A}$ of $S$. Transform each subexpression $SubRE$ into set of nodes with function $SubRE2Nodes(SubRE)$, then contract the SOA $\mathcal{A}$ with function contract($U, SubRE$). After this procedure we actually turn the SOA into a generalized SOA. The steps are shown in lines 9-11.
  \item 
  Finally we calculate level number for every nodes of generalized SOA and find all skip levels, if there are more than one $\&$ node (node with $\&$ operator) with the same level number, or if a level is a skip level, then ? is appended to every chain factor on that level. The lines 12-20 show the process and return an ISIRE in the end.
\end{enumerate}

\begin{algorithm}[!h]
  \caption{InferISIRE}
  \label{alg1}
     \KwIn{A set of strings $S$}
    \KwOut{An ISIRE}

             $\Sigma \leftarrow alphabet(S)$; $dictOper \leftarrow cntOper(S)$;\\
             $consist\_tr=CS(S), constraint\_tr \leftarrow NCS(S)$;\\
             $G \leftarrow Graph(\Sigma, constraint\_tr)$;\\
             $subGraphs \leftarrow connected\_subgraph(G)$;\\
             $R \leftarrow \emptyset$; $result \leftarrow \varepsilon$\\
             \ForEach{$subgraph$ in $subGraphs$}{
             $R.append(G2SubRE(subgraph,consist\_tr,dictOper))$;\\
             }
             Construct the $\mathcal{A} \leftarrow SOA(S)$ using \textit{2T}-\textit{INF} \cite{Garcia2002Inference};\\

             \ForEach{$SubRE$ in $R$}{
             $U \leftarrow SubRE2Nodes(SubRE)$;\\
              $\mathcal{A} \leftarrow A.contract(U, SubRE)$;\\
             }
             $\mathcal{A}.constructLevelOrder()$;\\
             \For{ $i = 1$ to $ln(\mathcal{A}.snk) - 1$}{
             $B \leftarrow$ all nodes with level number $i$ and $\&$;\\
             \ForEach{$\alpha$ in $B$}{
             \If{$\mathcal{A}.isSkipLevel(i)$ or $|B|>1$}{
             $result \leftarrow result \cdot \alpha^{?}$;\\
             }
             \lElse{$result \leftarrow result \cdot \alpha$;\\}
             }
             }
    \textbf{return} $result$
\end{algorithm}
\begin{algorithm}[!h]
  \caption{G2SubRE}
  \KwIn{An undirected graph $g$, a list $consist\_tr$, a map $dictOper$}
    \KwOut{A regular expression $r$}
    $allmis \leftarrow \emptyset$, $T \leftarrow \emptyset$;\\
    \While{$g.nodes() > 0$}{
            $mis \leftarrow clique\_removal(g)$;\\
            $g \leftarrow g- mis$;\\
			$allmis.append(mis)$;\\
    }
            \ForEach{$mis \in allmis$}{
            $dg \leftarrow Digraph(mis, Filter(mis, consist\_tr))$;\\
            $T.append(topological\_sort(dg))$;\\
         }
         $r \leftarrow learner_{oper}(dictOper,T)$;\\
    \textbf{return} $r$
\end{algorithm}

Here we give a brief explanation 
of the algorithm G2SubRE, which transforms an undirected graph to a subexpression. For an undirected graph $g$, the algorithm finds the approximation of maximum independent set (MIS) of $g$ with function $clique\_removal(g)$ \cite{DBLP:journals/bit/BoppanaH92}, then adds it to list $allmis$ and deletes the MIS and their related edges from $g$. The process is repeated until there exist no nodes in $g$.
For each MIS $mis$ in $allmis$, we establish a directed graph $dg$ with its filtered subset Filter($mis,consist\_tr$), then we compute the topological sort for all subgraphs of $dg$ and add the result to $T$. Finally, the algorithm returns the subexpression whose corresponding counting operators can be read from $dictOper$.

\textbf{Algorithm Analysis}. Let $m$ denotes the sum length of the input strings in $S$,  and $n$ denotes the number of alphabet symbols. It takes $O(m+n)$ to calculate all partial orders as well as constructing a graph. For a graph $G(V,E)$ with $|V|=n$ and $|E|=m$, it costs time $O(m+n)$ to find all maximal connected subgraphs. The time complexity of $clique\_removal()$ is $O(n^2+m)$. Thus for each subgraphs, computation of $allmis$ costs time $O(n^3+m)$. Since the number of maximal connected subgraphs of a graph is finite, the computation of $allmis$ for all subgraphs also costs time $O(n^3+m)$. Constructing SOA for $S$ and generalizing SOA can be finished in time $O(m+n)$, and assigning level numbers and computing all skip levels will be finished in time $O(m+n)$. All nodes will be converted into specific chain factors of \textit{ISIRE} in $O(n)$. Therefore, the time complexity of InferISIRE is $O(n^3+m)$.
\subsection{A Learning Example}
\textbf{The Language Size}. The Language Size (LS for short), which measures for the simplest deterministic expression that ``best'' describes sample $S$. Formally, the Language Size of regular expression is calculated as follows: $\mathcal{LS}(E)=\sum ^{n}_{i=0}|L(E)^{=i}|$ where $n = 2|\Sigma|+1$ and $|\Sigma|$ is the size of alphabet of expression $E$, 
and $|L(E)^{=i}|$ denote the number of words in language $L(E)$ of length $i$. Then the regular expression with the smaller value of $\mathcal{LS}(E)$ overgeneralizes $S$ the less.

\textbf{The Minimum Description Length measure}. To evaluate the preciseness of different learning algorithms,
we employ the data encoding cost proposed in \cite{DBLP:conf/isit/AdriaansV07}, which reflects the degree of generalization of a regular expression for a given set of samples $S$. The data encoding cost compares the size of $S$ with the size of the language defined by an inferred expression $E$,
$datacost(E,S)= \sum^{n}_{i=0}(2*log_2 i + log_2(^{|L^{=i}(E)|}_{|S^{=i}|}))$,
where $n = 2|\Sigma|+1$ as before, 
and ${L}^{=i}(E)$ is the subset of words in ${L}(E)$ that have length $i$. Suppose $E_1$ and $E_2$ are two generalizations of $S$. If $datacost(E_1, S) < datacost(E_2, S)$, then we say $E_1$ describes $S$ better which overgeneralizes $S$ less.

Let $S= \{aabcde,acdcfe,dbbcfe,adbcef\}$, we get $\Sigma=\{a,b,c,d,e,f\}$, $dictOper$ $=\{a:*; b:*;c:+;d:1;e:1;f:?\}$, $CS(S)=\{<c,d>,<d,c>,<b,d>,<d,b>,<f,e>,<e,f>\}$, $NCS(S)=\{<c,f>,<a,c>,<a,e>,<b,f>,<b,c>,<d,e>,<a,d>,<a,f>,<c,e>,<b,e>,<a,b>,<d,f>\}$. By recognizing alphabet symbols and $CS(S)$ we get the undirected graph $G$ showed in Fig. \ref{undir_gh}. We get three maximal connected subgraphs of $G$ using function $connected\_subgraph(G)$, namely, $G_1=(V_1=\{a\},E_1=\{\})$, $G_2=(V_2=\{e,f\},E_2=\{(e,f)\})$, $G_3=(V_3=\{b,c,d\},E_3=\{(b,d),(d,c)\})$, and they are shown in Fig. \ref{sub_gh} with different colors. Then we use G2SubRE to transform $G_1,G_2,G_3$ to subexpressions $r_1,r_2,r_3$. Here we introduce the transformation process of $G_3$. For graph $G_3$, we obtain MIS set $allmis$ = $\{\{b,c\},\{d\}\}$. Next, compute the topological sort for each MIS. Filter(\{b,c\},NCS(S))=$\{<b,c>\}$, so we add $bc$ to $T$. For $\{d\}$ there is only one node add $d$ to $T$, at last $T = \{bc, d\}$. Then using $learner_{oper}(dictOper,T)$, we concatenate all elements of $T$ with $\&$ and add unary operate for each symbol, and get the subexpression $r_3=b^*c^+\&d$. Simiarily $r_1=a^*$, $r_2=e\&f^?$. We construct the SOA $\mathcal{A}$ using algorithm \textit{2T}-\textit{INF} \cite{Garcia2002Inference}, the SOA $\mathcal{A}$ is shown in Fig. \ref{SOA_gh}. Then we contract SOA with $r_1,r_2,r_3$ and get generalized SOA presented in Fig. \ref{GSOA_gh}. We calculated level number for every nodes of generalized SOA and get the final inferred $result=a^*(b^*c^+\&d)(e\&f^?)$.

For this sample $S$, Table. \ref{table_res} shows the learning result of $learner^+_{DME}$ (for DME)\cite{DBLP:journals/corr/CiucanuS13}, conMiner (for SIRE) \cite{peng2015discovering} and InferISIRE. The learning result of $learner^+_{DME}$ allows all symbols to appear in any order, but in the sample $S$, from $NCS(S)$ we can get all ordered pairs of symbols, i.e., symbol $c$ always occurs before symbol $f$. Besides, in $S$, for all strings that symbol $a$ occurs, $a$ always occurs in the first position. Result of InferISIRE is consistent with this constrains while neither $learner^+_{DME}$ nor conMiner conform to this. Among the three algorithms, the result of InferISIRE
has the smallest Language Size and $datacost$, which shows preciseness of the regular expressions inferred by InferISIRE.
\begin{table}[ht]
  \caption[justification=centering]{The Learning Results of Different Algorithms}
  \label{table_res}
  \centering
 \scriptsize{
  \begin{tabular}{|c|c|c|c|}
     \hline
     Learning Method& Learning Result  & LS&datacost   \\
     \hline
     $learner^+_{DME}$ & $a^*\&b^*\&c^+\&d\&e\&f^?$ &$1.71*10^8$& $108.67$   \\
      \hline
     conMiner& $a^*de\&b^*c^+f^?$   &$1.79*10^5$& $92.45$   \\
      \hline
     InferISIRE& $a^*(b^*c^+\&d)(e\&f^?)$ & $4.86*10^3$& $85.66$ \\
     \hline
   \end{tabular}
  }
\end{table}

\begin{figure}
\begin{minipage}[b]{0.24\textwidth}
\scalebox{0.65}{
 \begin{tikzpicture}[shorten >=1pt,node distance=2cm,on grid,auto]
\node[state] (a) at (0,0) {$a$};
\node[state] (e) at (1.5,0) {$e$};
\node[state] (f) at (3,0) {$f$};
\node[state] (b) at (0,-1.5) {$b$};
\node[state] (c) at (3,-1.5) {$c$};
\node[state] (d) at (1.5,-1.5) {$d$};
\draw[liyt_gray] (b) -- (d) -- (c);
\draw[liyt_gray] (e) --(f);
\end{tikzpicture}
}
\caption{\scriptsize{Graph on $\Sigma$ \protect \\ and $CS(S)$}}
\label{undir_gh}
\end{minipage}
\begin{minipage}[b]{0.24\textwidth}
\scalebox{0.65}{
\begin{tikzpicture}[shorten >=1pt,node distance=2cm,on grid,>=stealth,thick,
every state/.style={fill,draw=none,orange,text=white,circular drop shadow},
a1/.style ={liyt_red,text=white},
a2/.style ={liyt_green,text=white},
a3/.style ={liyt_blue,text=white}]
\node[state,a1] (a) at (0,0) {$a$};
\node[state,a3] (e) at (1.5,0) {$e$};
\node[state,a3] (f) at (3,0) {$f$};
\node[state,a2] (b) at (0,-1.5) {$b$};
\node[state,a2] (c) at (3,-1.5) {$c$};
\node[state,a2] (d) at (1.5,-1.5) {$d$};
\draw[liyt_gray] (b) -- (d) -- (c);
\draw[liyt_gray] (e) --(f);
\end{tikzpicture}
}
\caption{\scriptsize{Three maximal connected subgraphs}}
\label{sub_gh}
\end{minipage}
\begin{minipage}[b]{0.24\textwidth}
\scalebox{0.65}{
 \begin{tikzpicture}[shorten >=1pt,node distance=1.3cm,on grid,auto,initial text=, state/.style={rectangle,draw=black!60},]
\node[state] (src) at (0,0)    {$src$};
\node[state] (a) [right =of src] {$a$};
\node[state] (b) [right =of a] {$b$};
\node[state] (d) [right =of b] {$d$};
\node[state,accepting](snk) [below =of src] {$snk$};
\node[state] (f) [right =of snk] {$f$};
\node[state] (c) [right =of f] {$c$};
\node[state] (e) [right =of c] {$e$};

 \path[->]
    (src) edge   (a)
    (src) edge [bend left = 37]  (d)
    (a) edge   (b)
    (a) edge   (c)
    (a) edge [bend left = 40]  (d)
        edge [loop above]  ()
    (b) edge   (c)
        edge [loop above]  ()
    (c) edge   (d)
    (c) edge   (e)
    (c) edge    (f)
    (d) edge   (b)
    (d) edge   (c)
    (d) edge   (e)
    (e) edge [bend left = 25]  (f)
    (e) edge [bend left = 30]   (snk)
    (f) edge [bend right = 25]  (e)
    (f) edge   (snk);

\end{tikzpicture}
}
\caption{\scriptsize{SOA of sample}}
\label{SOA_gh}
\end{minipage}
\begin{minipage}[b]{0.25\textwidth}
\scalebox{0.65}{
 \begin{tikzpicture}[shorten >=1pt,node distance=1.3cm,on grid,auto,initial text=, state/.style={rectangle,draw=black!60},
 a1/.style ={fill,liyt_red,text=white},
a2/.style ={fill,liyt_green,text=white},
a3/.style ={fill,liyt_blue,text=white}]
\node[state] (src) at (0,0)    {$src$};
\node[state,a1] (a) [right = of src] {$a^*$};
\node[state,a2] (b) [below = of a] {$b^{*}c^{+}\&d$};
\node[state,a3] (f) [right  = of a] {$e\&f^{?}$};
\node[state,accepting](snk) [right = of b] {$snk$};
 \path[->]
    (src) edge   (a)
    (src) edge   (b)
    (a) edge   (b)
    (b) edge    (f)
    (f) edge   (snk);

\end{tikzpicture}
}
\caption{\scriptsize{Generalized SOA after contracting}}
\label{GSOA_gh}
\end{minipage}
\end{figure}

\section{Experiments}
In this section, we first analyse the usage proportion of ISIREs and other subclasses based on large-scale real data.
Then we compare our learning algorithm InferISIRE with other learning algorithms of different subclasses ($Learn_{DME}^+$ for DME \cite{DBLP:journals/corr/CiucanuS13}, conMiner for SIRE \cite{peng2015discovering})) and one XML tool (InstanceToSchema \cite{InstanceToSchema}). All experiments were conducted on a machine with Intel Core i5-5200U@2.20GHz, 4G memory. All codes are written in python 3.
\subsection{Experiment of Practicality Analysis}
\begin{wrapfigure}[10]{r}{0.45\textwidth}
\captionsetup{justification=centering}
\vspace{-25pt}
\includegraphics[width=1.9in,height=1.3in]{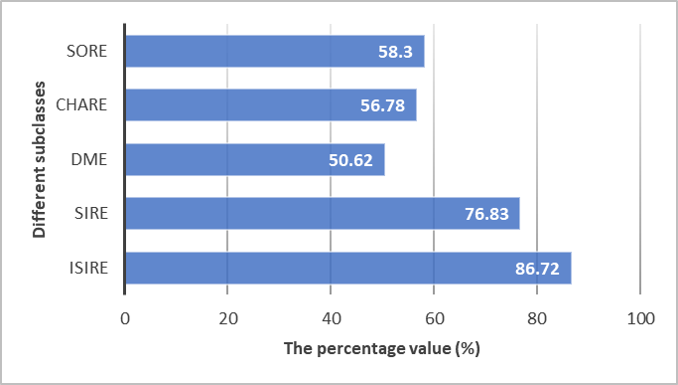}
\captionsetup{font=scriptsize}
\captionof{figure}{Proportions of subclasses}
\label{pro_fig}
\end{wrapfigure}
We crawled 4,872 distinct Relax NG schema files from 254 websites, and 103 Github repositories utilizing Google Search Engine. We extracted 137,286 regular expressions from these files, and found that 38.45\% expressions have the interleaving
operator, which shows its widespread use in practice. Based on this 137,286 regular expressions, we investigated the usage proportion of ISIREs compared with other popular subclasses. The experimental result is shown in Fig.\ref{pro_fig}. Note that SORE \cite{Bex2010Inference} and CHARE \cite{Freydenberger2015Fast} are subclasses of standard regular expressions, and DME \cite{DBLP:conf/webdb/BonevaCS13,DBLP:journals/corr/CiucanuS13}, SIRE \cite{peng2015discovering} and ISIRE are subclasses supporting \&.
It is obvious from Fig. \ref{pro_fig} that ISIRE has the highest proposition. The result shows ISIREs are more practical in real-world applications.

\subsection{Analysis on Learning Results of Different Algorithms}
We downloaded three XML data corpora, including DBLP (computer science bibliography corpus), NASA (Datasets converted from legacy flat-file format into XML), DocBook (a schema maintained by the DocBook Technical Committee of OASIS). Based on sets of strings extracted from these corpora, we compare the learning results of InferISIRE, the XML tool InstanceToSchema \cite{InstanceToSchema} (i2s for short), $learner^+_{DME}$ \cite{DBLP:journals/corr/CiucanuS13} and conMiner \cite{peng2015discovering}, which all support learning a regular expression with interleaving from XML data.

We first use the Language Size and $datacost$ to measure the preciseness of expressions for a set of samples. However, during experiments on real data, we found that when the alphabet is large, the calculation of Language Size and $datacost$ becomes extraordinarily time-consuming, since it needs a lot of tedious work for computing the number of subset words of $L(E)$. Given an expression $E =a^*b^?(c^*\&d^?)ef^+$ as an example, $n=2*6+1=13$, actually it is tough to compute the number of words in $L(E)$ with length 13. Thus we introduce a simple definition called Combinatorial Cardinality to value the precision of expressions. Combinatorial Cardinality is limited to measure the precision of expressions with interleaving, satisfying that different expressions have the same symbols with the same unary operators $\{1,?,*,+\}$. The smaller value of Combinatorial Cardinality, the more precise of an expression. It is defined as follows.
\begin{definition}
\textbf{Combinatorial Cardinality ($CC$)} \cite{LiZXMC18} Let $\Sigma$ be the finite alphabet set. $E_i$ is a regular expression with interleaving over $\Sigma$. $a,b$$\in$$\Sigma$ and $u,u',v,v'$$\in$$\Sigma^*$.
\begin{itemize}
  \item $CC(a)=CC(\varepsilon)=1$, where $a\in \Sigma$;
  \item $CC(E^t)=CC(E)$, where $t\in rep$;
  \item $CC(E_1|E_2\cdots | E_n)=CC(E_1)+CC(E_2)+\cdots +CC(E_n)$;
  \item $CC(E_1\cdot E_2\cdots E_n)=CC(E_1)\times CC(E_2)\times \cdots \times CC(E_n)$;
  \item $CC(u\& v)=CC(a\cdot (u'\&v))+CC(b\cdot (u\&v'))$, where $u=au'$ and $v=bv'$;
  \item $CC((E_1 | E_2 | \cdots | E_m)\& (E_1' | E_2' | \cdots | E_n'))=CC(\bigcup_{i=1\\j=1}^{i=m\\j=n}E_i\&E_j')$.
\end{itemize}
\end{definition}

\begin{figure}
\begin{minipage}{0.48\textwidth}
            \centering
            \scriptsize
            \makeatletter\def\@captype{table}\makeatother\caption{The Results on NASA}
            \label{NASA}
            \scriptsize
            \renewcommand\arraystretch{1.4}
            \begin{tabular}{|c|c|c|c|}

\hline
{Method}& {LS} &{datacost}& {CC}  \\

\hline
i2s& $8.66*10^{8}$ &$1.95*10^3$&  $3.63*10^{6}$    \\
\hline
learner$^+_{DME}$ &$8.66*10^{8}$ &$1.95*10^3$&  $3.63*10^{6}$    \\
\hline
conMiner& $1.42*10^{8}$ &$1.67*10^3$&  $10$    \\
\hline
InferISIRE &$2.60*10^{5}$ &$1.00*10^3$&  $2$    \\
\hline
\end{tabular}
\end{minipage}
\begin{minipage}{0.48\textwidth}
        \centering
        \scriptsize
        \makeatletter\def\@captype{table}\makeatother\caption{The Results on www(DBLP)}
        \label{wwwlabel}
        \scriptsize
        \renewcommand\arraystretch{1.4}
        \begin{tabular}{|c|c|c|c|}
\hline
{Method}& {LS} &{datacost}& {CC}  \\

\hline
i2s& $1.53*10^{18}$ &$1.34*10^4$&  $3.63*10^{6}$    \\
\hline
learner$^+_{DME}$ &$1.43*10^{15}$ &$1.11*10^4$&$2160$    \\
\hline
conMiner& $5.71*10^{12}$ &$9.50*10^3$&$360$    \\
\hline
InferISIRE &$2.35*10^{10}$ &$6.47*10^3$&$168$   \\
\hline
\end{tabular}
\end{minipage}

\begin{minipage}{0.48\textwidth}
            \centering
            \scriptsize
            \makeatletter\def\@captype{table}\makeatother\caption{The Results on phdthesis(DBLP)}
            \label{phdthesis}
            \scriptsize
            \renewcommand\arraystretch{1.4}
            \begin{tabular}{|c|c|c|c|}

\hline
{Method}& {LS} &{datacost}& {CC}  \\

\hline
i2s& $5.89*10^{25}$ &$5.63*10^3$&  $8.72*10^{10}$    \\
\hline
learner$^+_{DME}$ &$7.04*10^{24}$ &$5.27*10^3$&$3.27*10^{7}$    \\
\hline
conMiner& $1.24*10^{18}$ &$4.28*10^3$&$1.51*10^{7}$    \\
\hline
InferISIRE &$4.07*10^{14}$ &$4.04*10^3$&$8.32*10^{5}$   \\
\hline
\end{tabular}
\end{minipage}
\begin{minipage}{0.48\textwidth}
        \centering
        \scriptsize
        \makeatletter\def\@captype{table}\makeatother\caption{The Results on DocBook}
        \label{DocBook}
        \scriptsize
        \renewcommand\arraystretch{1.4}
            \begin{tabular}{|c|c|c|c|}

\hline
{Method}& {LS} &{datacost}& {CC}  \\

\hline
i2s& $7.50*10^{12}$ &$3.75*10^6$&  $4.03*10^{4}$    \\
\hline
learner$^+_{DME}$ &$3.57*10^{11}$ &$3.19*10^6$&$1.01*10^{4}$    \\
\hline
conMiner& $8.43*10^{7}$ &$1.50*10^6$&$56$    \\
\hline
InferISIRE &$2.69*10^{6}$ &$7.81*10^5$&$6$   \\
\hline
\end{tabular}
\end{minipage}
\end{figure}

\begin{table*}[h]
\centering
\renewcommand\arraystretch{1.3}
\caption{Results of Inference Using Different Methods on XML datasets}
\label{big_exp}
\scalebox{0.85}{
\begin{tabular}{|c|c|c|}
\hline
\textbf{Dataset}&\textbf{Result of InstanceToSchema}\\
\textbf{Name}&\textbf{Result of learner$^+_{DME}$}\\
\textbf{Sample}&\textbf{Result of conMiner}\\
\textbf{Size}&\textbf{Result of InferISIRE}\\

\hline
&$a_{1}\&a_{2}^{+}\&a_{3}\&a_{4}^{+}\&a_{5}\&a_{6}^{*}\&a_{7}^{*}\&a_{8}\&a_{9}\&a_{10}^{*}$\\

\textbf{NASA}&$a_{1}\&a_{2}^{+}\&a_{3}\&a_{4}^{+}\&a_{5}\&a_{6}^{*}\&a_{7}^{*}\&a_{8}\&a_{9}\&a_{10}^{*}$\\

2435&$a_{1}a_{2}^{+}a_{3}a_{4}^{+}a_{5}a_{6}^{*}a_{7}^{*}a_{8}a_{9}\&a_{10}^{*}$\\

&$a_{1}a_{2}^{+}a_{3}a_{4}^{+}a_{5}a_{6}^{*}(a_{10}^{*}\&a_{7}^{*})a_{8}a_{9}$\\
\hline

&$a_{1}\&a_{2}\&a_{3}^{?}\&a_{4}^{*}\&a_{5}^{*}\&a_{7}^{*}\&a_{6}^{*}\&a_{8}^{?}$\\

\textbf{table(DocBook)}&$(a_{6}^{*}|a_{7}^{*})\&a_{5}^{*}\&a_{4}^{*}\&a_{1}\&a_{2}\&a_{3}^{?}\&a_{8}^{?}$\\

1719728&$a_{3}^{?}\&a_{4}^{*}\&a_{1}a_{2}a_{5}^{*}a_{7}^{*}a_{6}^{*}a_{8}^{?}$\\

&$a_{1}a_{2}(a_{3}^{?}\&a_{4}^{*}\&a_{5}^{*})a_{7}^{*}a_{6}^{*}a_{8}^{?}$\\

\hline

&$a_{7}^{?}\&a_{10}^{*}\&a_{1}^{*}\&a_{17}^{?}\&a_{2}^{*}\&a_{5}^{*}\&a_{14}^{*}\&a_{3}^{?}\&a_{6}^{?}\&a_{15}^{*}$\\

\textbf{www(DBLP)}&$(a_{10}^{*}|a_{17}^{?}|a_{6}^{?})\&(a_{5}^{*}|a_{3}^{?}|a_{7}^{?})\&(a_{2}^{*}|a_{1}^{*})\&a_{14}^{*}\&a_{15}^{*}$\\

2000226&$a_{6}^{?}a_{1}^{*}a_{14}^{*}a_{5}^{*}a_{3}^{?}a_{7}^{?}a_{17}^{?}\&a_{15}^{*}\&a_{2}^{*}a_{10}^{*}$\\

&$a_{1}^{*}a_{6}^{?}(a_{14}^{*}a_{5}^{*}\&a_{15}^{*}\&a_{2}^{*}a_{10}^{*}a_{3}^{?}a_{7}^{?}a_{17}^{?})$\\

\hline
&$a_{7}^{*}\&a_{10}^{*}\&a_{17}^{+}\&a_{2}^{+}\&a_{19}^{*}\&a_{14}\&a_{15}^{?}\&a_{16}^{?}\&a_{11}^{?}\&a_{12}^{*}\&a_{9}^{?}\&a_{21}^{*}\&a_{20}^{?}\&a_{13}^{?}$\\

\textbf{phdthesis(DBLP)}&$(a_{11}^{?}|a_{9}^{?}|a_{19}^{*})\&(a_{10}^{*}|a_{20}^{?}|a_{15}^{?})\&a_{7}^{*}\&a_{16}^{?}\&a_{12}^{*}\&a_{13}^{?}\&a_{2}^{+}\&a_{14}\&a_{21}^{*}\&a_{17}^{+}$\\

63420&$a_{13}^{?}a_{20}^{?}\&a_{12}^{*}a_{7}^{*}\&a_{17}^{+}a_{16}^{?}\&a_{2}^{+}a_{14}a_{9}^{?}a_{21}^{*}a_{15}^{?}\&a_{11}^{?}a_{19}^{*}a_{10}^{*}$\\

&$a_{2}^{+}a_{14}(a_{19}^{*}\&a_{13}^{?}a_{20}^{?}\&a_{17}^{+}a_{16}^{?}a_{11}^{?}\&a_{21}^{*}a_{10}^{*}\&a_{12}^{*}a_{9}^{?}a_{15}^{?}a_{7}^{*})$\\

\hline
\end{tabular}
}
\label{article}
\end{table*}

The inferred regular expressions on these three XML corpora are shown in Table. \ref{big_exp}. For every XML data set, the inferred regular expressions of InstanceToSchema \cite{InstanceToSchema}, conMiner \cite{peng2015discovering}, $learner^+_{DME}$ \cite{DBLP:journals/corr/CiucanuS13} and InferISIRE are shown from top to bottom. To save space, we use the short names of words and the list of abbreviations is shown in \url{https://github.com/yetingli/sofsem2019}. We calculated the Language Size, $datacost$ and $CC$ for the inferred results of different methods on each data set, and they are presented in Table. \ref{NASA}-\ref{DocBook} respectively. We find that all regular expressions inferred by InstanceToSchema are only symbols combined with the interleaving, not surprisingly, its result has a maximum Language Size, $datacost$ and $CC$ on each data set, which denotes overgeneralization of InstanceToSchema. Meanwhile, we find that the Language Size, $datacost$ and $CC$ values of regular expressions inferred by InferISIRE are the smallest on all the data sets, which indicates its better preciseness and the least generalization among these methods.

Taking the results on NASA as an example. Both InstanceToSchema and $learner^+_{DME}$ only use interleaving to connect all symbols, thus some sequential restrictions have been lost and learning results are overgeneralization.
Furthermore, conMiner can not learn some consistent partial orders due to the limits in SIREs, that is the \& operator can only appears at the outermost layer of an expression in SIRE. In the strings set extracted from nasa.xml, $a_1$ always occurs at the first position of strings and $a_9$ always occurs at the last position of strings, the result of InferISIRE conforms to this constrain. 
But the result of conMiner allows $a_{10}$ to appear before $a_1$ or after $a_9$, which makes the result overgeneralization for samples. To conclude, the regular expressions inferred by InferISIRE are more precise compared with other tools and methods.

\section{Conclusion}
In this paper, focusing on the inference of Relax NGs, we proposed a new subclass ISIREs of regular expressions with interleaving. Then we designed a polynomial inference algorithm InferISIRE to learn ISIREs. Based on large-scale real data, we calculated the usage proportion of ISIREs and other popular subclasses, and found that the proportion of ISIREs is the highest, which shows ISIREs have a better practicality. At last we compare the inferred results of InferISIRE with other inference algorithms on three XML data corpora, experimental results showed that the regular expressions inferred by InferISIRE are more precise.

\bibliography{lipics-v2016-sample-article}

\begin{thebibliography}{10}

\bibitem{abiteboul2015highly}
Serge Abiteboul, Pierre Bourhis, and Victor Vianu.
\newblock {Highly Expressive Query Languages for Unordered Data Trees}.
\newblock {\em J. {Theory Comput. Syst}}, 57(4):927--966, 2015.

\bibitem{Abiteboul2000Data}
Serge Abiteboul, Peter Buneman, and Dan Suciu.
\newblock {\em Data on the Web: From Relations to Semistructured Data and
  {XML}}.
\newblock Morgan Kaufmann, 1999.

\bibitem{DBLP:conf/isit/AdriaansV07}
Pieter~W. Adriaans and Paul M.~B. Vit{\'{a}}nyi.
\newblock The power and perils of {MDL}.
\newblock In {\em Proc. 12. {ISIT}}, pages 2216--2220, 2007.

\bibitem{Benedikt2008XPath}
Michael Benedikt, Wenfei Fan, and Floris Geerts.
\newblock {XPath Satisfiability in the Presence of DTDs}.
\newblock {\em J. {ACM}}, 55(2):1--79, 2008.

\bibitem{Bex2006Inference}
Geert~Jan Bex, Frank Neven, Thomas Schwentick, and Karl Tuyls.
\newblock {Inference of Concise DTDs from XML Data}.
\newblock In {\em Proc. 32. {VLDB}}, pages 115--126, 2006.

\bibitem{Bex2010Inference}
Geert~Jan Bex, Frank Neven, Thomas Schwentick, and Stijn Vansummeren.
\newblock {Inference of Concise Regular Expressions and DTDs}.
\newblock {\em J. {ACM Trans. Database Syst}}, 35(2):1--47, 2010.

\bibitem{Bex2007Inferring}
Geert~Jan Bex, Frank Neven, and Stijn Vansummeren.
\newblock {Inferring XML Schema Definitions from XML Data}.
\newblock In {\em Proc. 33. {VLDB}}, pages 998--1009, 2007.

\bibitem{DBLP:conf/webdb/BonevaCS13}
Iovka Boneva, Radu Ciucanu, and Slawek Staworko.
\newblock {Simple Schemas for Unordered XML}.
\newblock In {\em Proc. 13. {WebDB}}, pages 13--18, 2013.

\bibitem{DBLP:journals/corr/BonevaGPS15}
Iovka Boneva, Jos{\'{e}} Emilio~Labra Gayo, Eric~G. Prud'hommeaux, and Slawek
  Staworko.
\newblock {Shape Expressions Schemas}.
\newblock {\em CoRR}, abs/1510.05555, 2015.

\bibitem{DBLP:journals/bit/BoppanaH92}
Ravi~B. Boppana and Magn{\'{u}}s~M. Halld{\'{o}}rsson.
\newblock {Approximating Maximum Independent Sets by Excluding Subgraphs}.
\newblock {\em J. {BIT}}, 32(2):180--196, 1992.

\bibitem{Che2006Query}
Dunren Che, Karl Aberer, and M.~Tamer \"{O}zsu.
\newblock {Query Optimization in XML Structured-Document Databases}.
\newblock {\em J. {VLDB}}, 15(3):263--289, 2006.

\bibitem{DBLP:journals/corr/CiucanuS13}
Radu Ciucanu and Slawek Staworko.
\newblock {Learning Schemas for Unordered XML}.
\newblock In {\em Proc. 14. {DBPL}}, 2013.

\bibitem{InstanceToSchema}
Didier Demany.
\newblock {InstanceToSchema: a Relax NG Schema Generator from XML Instances}.
\newblock {\em http://www.xmloperator.net/i2s/}, 2005.

\bibitem{DBLP:conf/kdd/FowkesS16}
Jaroslav~M. Fowkes and Charles~A. Sutton.
\newblock {A Subsequence Interleaving Model for Sequential Pattern Mining}.
\newblock In {\em Proc. 22. {KDD}}, pages 835--844, 2016.

\bibitem{Freydenberger2015Fast}
Dominik~D Freydenberger and Timo K\"{o}tzing.
\newblock {Fast Learning of Restricted Regular Expressions and DTDs}.
\newblock {\em J. {Theory Comput. Syst}}, 57(4):1114--1158, 2015.

\bibitem{Garcia2002Inference}
P.~Garcia and E.~Vidal.
\newblock {Inference of k-Testable Languages in the Strict Sense and
  Application to Syntactic Pattern Recognition}.
\newblock {\em J. {IEEE Trans. Pattern Anal. Mach. Intell}}, 12(9):920--925,
  2002.

\bibitem{DBLP:journals/datamine/GarofalakisGRSS03}
Minos~N. Garofalakis, Aristides Gionis, Rajeev Rastogi, S.~Seshadri, and
  Kyuseok Shim.
\newblock {XTRACT: Learning Document Type Descriptors from XML Document
  Collections}.
\newblock {\em J. {Data Min. Knowl. Discov}}, 7(1):23--56, 2003.

\bibitem{Gelade2008Succinctness}
Wouter Gelade.
\newblock {Succinctness of Regular Expressions with Interleaving, Intersection
  and Counting}.
\newblock In {\em Proc. 33. {MFCS}}, pages 363--374, 2008.

\bibitem{DBLP:conf/dbpl/GhelliCS07}
Giorgio Ghelli, Dario Colazzo, and Carlo Sartiani.
\newblock {Efficient Inclusion for a Class of {XML} Types with Interleaving and
  Counting}.
\newblock In {\em Proc. 11. {DBPL}}, pages 231--245, 2007.

\bibitem{Gold1967Language}
E~Mark Gold.
\newblock {Language Identification in the Limit}.
\newblock {\em J. {Information and Control}}, 10(5):447--474, 1967.

\bibitem{grijzenhout2013quality}
Steven Grijzenhout and Maarten Marx.
\newblock {The Quality of the XML Web}.
\newblock {\em J. {Web Sem}}, 19:59--68, 2013.

\bibitem{DBLP:conf/lata/KimKH16}
Guen{-}Hae Kim, Sang{-}Ki Ko, and Yo{-}Sub Han.
\newblock {Inferring a Relax {NG} Schema from {XML} Documents}.
\newblock In {\em Proc. 10. {LATA}}, pages 400--411, 2016.

\bibitem{DBLP:conf/chi/KinHDA12}
Kenrick Kin, Bj{\"{o}}rn Hartmann, Tony DeRose, and Maneesh Agrawala.
\newblock {Proton: Multitouch Gestures as Regular Expressions}.
\newblock In {\em {CHI} Conference on Human Factors in Computing Systems},
  pages 2885--2894, 2012.

\bibitem{Koch2004Schema}
Christoph Koch, Stefanie Scherzinger, Nicole Schweikardt, and Bernhard
  Stegmaier.
\newblock {Schema-Based Scheduling of Event Processors and Buffer Minimization
  for Queries on Structured Data Streams}.
\newblock In {\em Proc. 30. {VLDB}}, pages 228--239, 2004.

\bibitem{LiZXMC18}
Yeting Li, Xiaolan Zhang, Han Xu, Xiaoying Mou, and Haiming Chen.
\newblock Learning restricted regular expressions with interleaving from {XML}
  data.
\newblock In {\em Proc. 37. {ER}}, pages 586--593, 2018.

\bibitem{DBLP:conf/icde/LiG15}
Zheng Li and Tingjian Ge.
\newblock {PIE: Approximate Interleaving Event Matching over Sequences}.
\newblock In {\em Proc. 31. {ICDE}}, pages 747--758, 2015.

\bibitem{Manolescu2001Answering}
Ioana Manolescu, Daniela Florescu, and Donald Kossmann.
\newblock {Answering XML Queries on Heterogeneous Data Sources}.
\newblock In {\em Proc. 27. {VLDB}}, pages 241--250, 2001.

\bibitem{Martens2003Typechecking}
Wim Martens and Frank Neven.
\newblock {Typechecking Top-Down Uniform Unranked Tree Transducers}.
\newblock In {\em Proc. 9. {ICDT}}, pages 64--78, 2003.

\bibitem{Martens2004Frontiers}
Wim Martens and Frank Neven.
\newblock {Frontiers of Tractability for Typechecking Simple XML
  Transformations}.
\newblock In {\em Proc. 23. {PODS}}, pages 23--34, 2004.

\bibitem{Min2003Efficient}
Jun~Ki Min, Jae~Yong Ahn, and Chin~Wan Chung.
\newblock {Efficient Extraction of Schemas for XML Documents}.
\newblock {\em J. {Inf. Process. Lett}}, 85(1):7--12, 2003.

\bibitem{DBLP:journals/siamcomp/PaigeT87}
Robert Paige and Robert~Endre Tarjan.
\newblock {Three Partition Refinement Algorithms}.
\newblock {\em {SIAM} J. Comput.}, 16(6):973--989, 1987.

\bibitem{Papakonstantinou2000DTD}
Yannis Papakonstantinou and Victor Vianu.
\newblock {DTD Inference for Views of XML Data}.
\newblock In {\em Proc. 19. {PODS}}, pages 35--46, 2000.

\bibitem{peng2015discovering}
Feifei Peng and Haiming Chen.
\newblock {Discovering Restricted Regular Expressions with Interleaving}.
\newblock In {\em Proc. 17. {APWeb}}, pages 104--115. Springer, 2015.

\bibitem{DBLP:conf/icdt/StaworkoBGHPS15}
Slawek Staworko, Iovka Boneva, Jos{\'{e}} Emilio~Labra Gayo, Samuel Hym,
  Eric~G. Prud'hommeaux, and Harold~R. Solbrig.
\newblock {Complexity and Expressiveness of ShEx for RDF}.
\newblock In {\em Proc. 18. {ICDT}}, pages 195--211, 2015.

\end{thebibliography}
\end{document}